\documentclass[preprint,aps]{revtex4-2}

\begin{document}

\centerline{\bf\large New aspects of gauge-gravity relation}

\bigskip

\centerline{A.G. Shuvaev}

\medskip
\centerline{\em
NRC "Kurchatov Institute" - PNPI, Gatchina 188300 Russia
}
\vskip 0.9 truecm
\centerline{E-mail: shuvaev@thd.pnpi.spb.ru}

\bigskip

\centerline{\large Abstract}
{\small
The relation between four-dimensional $SO(4)$ pure Yang-Mills theory
and the gravity is discussed.
The functional integral for Yang-Mills theory is rewritten in terms of
the gravity metric and Riemann tensors.
This relation is shown to also provide
a simple way to derive the linear potential
resulting from the average Wilson loop in
pure Yang-Mills theory.
}

\bigskip

\section{Introduction}

There has been much recent theoretical activity pertaining to
the relations between gauge fields and gravity.
One first has to cite the AdS/CFT correspondence relating
quantum field theory to the classical dynamics of gravity
in one higher dimension \cite{Maldacena:1997re, Witten:1998qj,
Gubser:1998bc, Aharony:1999ti}.
Significant progress has also been made in the area
of low-dimensional theories.
A lot of studies addressed the relationship of three-dimensional (3D)
pure quantum gravity with Chern-Simons gauge theory~\cite{Witten:2007kt}.
3D gravity has no propagating degrees of freedom
in the bulk, and the dynamics arises at the boundary of spacetime
if it exists and is governed there by the two-dimensional (2D)
Wess-Zumino-Witten model; see Refs.~\cite{Kiran:2014dfa,
Donnay:2016iyk} and references therein.
There is a lot of progress in 2D gravity models,
particularly in studying Jackiw-Teitelboim gravity,
which has the merit of being solvable, renormalizible,
and admitting AdS$_2$ solutions;
see \cite{Mertens:2022irh, Grumiller:2016dbn} for a review.
Nowadays, the topological $BF$ theories are of great interest
as a researh subject relevant for alternative theories of gravity
and gauge theories in several dimensions;
see the reviews \cite{Baez:1999sr, Freidel:2012np, Celada:2016jdt}.

The starting point of the following treatment is based
on the first-order formalism of general relativity
\cite{Plebanski, Halpern, Krasnov}. Gravity is formulated
in terms of the spin connection rather than
the spacetime metric \cite{Capovilla}, identifying the spin
connection with the $SU(2)$ gauge field.
This formalism was used to rewrite
3D Yang-Mills (YM) theory through gauge-invariant variables
\cite{Lunev_1, Lunev_2, Ganor:1995em, Anishetty:1992xa,
Anishetty:1998xn, Diakonov:2001xg}
in a way that brings
the action of the gauge theory to a form close to the Einstein-Hilbert
one. Apart from being a step toward unification of both theories,
it could be of interest by itself
as a way to gain deeper insight into the nature of YM theory.

Here we attempt to extend this approach to four-dimensional (4D) YM theory.
The main idea goes as
follows.Let us take the action
$$
S_{B}\,\sim\,
\int d^{\,4}x\,\varepsilon^{\mu\nu\lambda\sigma}
G_{\mu\nu}^{AB}(A)B_{\lambda\sigma}^{AB},
$$
with the YM strength tensor $G_{\mu\nu}^{AB}(A)$.
In fact, it looks like the action of $BF$ theory.
If the dual field variables are constrained to the form
$B_{\mu\nu}^{AB}=e_\mu^A e_\nu^B - e_\nu^A e_\mu^B$
through the four-vectors $e_\mu^A$ it turns into
the Hilbert-Palatini action $S_{\mathrm{HP}}(A,e)$.
The Euclidean integral over the gluon field $A_\mu$ yields
$\int DA_\mu e^{-S_{\mathrm{HP}}}\sim e^{i S_{\mathrm{EH}}}$,
where $S_{\mathrm{EH}}$ is the standard Einstein-Hilbert gravity
action.
On the other hand by choosing an appropriate weight function
$\rho(e_\mu^A)$, one gets the YM action,
$\int De_\mu^A \rho(e_\mu^A)e^{-S_{\mathrm{HP}}}\sim e^{-S_{\mathrm{YM}}}$.

We elaborate on two aspects of these relations.
The first is the connection between YM theory and the gravity
demonstrated by the partition functions. Its peculiar feature is
the cosmological term added to the Einstein-Hilbert action.
The second aspect is probably more interesting.
It concerns the YM theory in itself regardless of gravity.
Remarkably, it looks like the absence of gravity helps to derive
the linear potential for the Wilson loop and, in a sense, causes it.
These issues are addressed in the Secs.~III and IV.
Section~II
details the evaluation of the integrals that the subsequent analysis is based on.

\section{Auxiliary integrals}

We treat 4D YM theory in Euclidean space with $SO(4)$
gauge group. The gauge fields are $4\times 4$ antisymmetric matrices,
$A_\mu^{AB}(x)= -A_\mu^{BA}(x)$,
$\mu=1,\ldots,4$, $A,B=1,\ldots,4$. The strength tensor reads
$$
G_{\mu\nu}^{AB}(A)\,=\,\biggl(\partial_\mu A_\nu
- \partial_\mu A_\nu + \bigl[A_\mu,A_\nu\bigr]\biggr)^{AB},
$$
and the action is
$$
S\,=\,\frac 1{g^2}\int d^{\,4}x\,
G_{\mu\nu}^{AB}(A)\,G_{\mu\nu}^{AB}(A).
$$

To begin, we show that
the gluon partition function
can be presented as the functional integral
\begin{equation}
\label{Ze}
Z\,=\,\int DA_\mu De_\mu^A\,\exp\int d^4x
\left[\,-M^4(e_\mu^A e_\mu^A)^2\,
+\,ik M^2\,F(A)\,\right],
\end{equation}
where $e_\mu^A(x)$ are the auxiliary fields,
\begin{eqnarray}
F(A)\,&=&\,\mathcal{ F}^{AB,\mu\nu}(A)\,\Sigma_{\mu\nu}^{AB}, \nonumber \\
\label{F}
\mathcal{ F}^{AB,\mu\nu}(A)\,&=&\,
\varepsilon^{\mu\nu\lambda\sigma}\varepsilon^{ABCD} G_{\lambda\sigma}^{AB}(A),~~~
\Sigma_{\mu\nu}^{AB}\,=\,e_\mu^A e_\nu^B\,-\, e_\nu^A e_\mu^B,
\end{eqnarray}
the mass factor $M$ is introduced for correct dimensionality,
and $k$ is a parameter related to the coupling constant.

Note that there are no derivatives of the $e_\mu^A(x)$ field
in the integrand. As the functional measure is
$De_\mu^A=\prod\limits_x de_\mu^A(x)$
the integral over the auxiliary fields turns into
the product of the separate integrals over all space points.
To make sense of this one has to pass to discrete space
by imposing a cubic grid, with the lattice spacing playing the role
of the UV cutoff. Choosing the parameter $a$
as a minimal space distance, we have with
$F(x)\equiv F\bigl(A(x)\bigr)$
\begin{equation}
\label{ZAdis}
Z[A]\,=\,\prod_{x}\int de_\mu^A(x)
\exp\sum_{x}\left[-M^4\bigl(e_\mu^A(x) e_\mu^A(x)\bigr)^2\,a^4
\,+\,ik M^2 F(x)a^4
\right],
\end{equation}
or, after rescaling $e_\mu^A \to e_\mu^A/M$,
\begin{equation}
\label{Zprod}
Z[A]\,=\,C_0\,\prod_{x}\int de_\mu^A(x)\exp
\sum_{x}\left[-\bigl(e_\mu^A(x)\,e_\mu^A(x)\bigr)^2
+\,i k F(x)a^2\right],
\end{equation}
with the constant factor $C_0=\prod_x(\mu a)^{-16}$.
The $e_\mu^A(x)$ integrals can
be done by expanding the $Z(A)$ integrand into $a$ powers,
\begin{equation}
\label{IntSum}
Z[A]\,=\,Z_0\,\prod_{x}\,\bigl[1 - \frac 89\,k^2
G^{\,2}(x)a^4 + \mathcal{ O}(a^4)\bigr]
\,=\,Z_0\,\exp\left[- \frac 89\,k^2
\sum_{x}G^{\,2}(x)a^4 + \mathcal{ O}(a^4)\right],
\end{equation}
$$
G^{\,2}(x) \equiv G_{\mu\nu}^{AB}(x)G_{\mu\nu}^{AB}(x),
$$
where the terms of $a^2$ order vanish because of the strength tensor
antisymmetry with respect to the color or space indices.
Going to the continuous limit, $a \to 0$, and recognizing
the integral sum on the right-hand side of the Eq.~(\ref{IntSum}),
we finally arrive at the desired relation of
the functional integral (\ref{Ze}) to the partition function
of the $SO(4)$ gauge field,
\begin{equation}
\label{pf}
Z\,=\,Z_0\,\int DA_\mu\,\exp\left[-\frac 1{g^2}\int d^{\,4}x\,
G_{\mu\nu}^{AB}(A)\,G_{\mu\nu}^{AB}(A)\right],
\end{equation}
with the coupling constant
\begin{equation}
\label{g2}
\frac 1{g^2}\,=\,\frac 89\,k^2.
\end{equation}

It is worth pointing out that the initial relation (\ref{Ze})
may be replaced with a more general one
\begin{equation}
\label{Zrho}
Z\,=\,\int DA_\mu De_\mu^A\,\rho(e_\mu^A)\,
\exp\left[\,ik M^2\!\int\, d^{\,4}x
\,F(A)\,\right],
\end{equation}
with a local density $\rho(e_\mu^A)$. The locality means
the density is put at each space point,
\begin{equation}
\label{Derho}
De_\mu^A\rho(e_\mu^A)\,=\,\prod\limits_x de_\mu^A(x)
\rho\bigl(e_\mu^A(x)\bigr),
\end{equation}
and that it has no derivatives inside.
Any local density would be suitable provided
the integral (\ref{Zrho}) returns the partition function (\ref{pf})
in the continuous limit $a \to 0$.
The density that occurs in the relation (\ref{Ze}) reads
\begin{equation}
\label{rho1}
\rho(e_\mu^A)\,=\,
\exp\bigl[-M^4\bigl(e_\mu^A(x) e_\mu^A(x)\bigr)^2\,a^4\bigr].
\end{equation}
Another example,
\begin{equation}
\label{rho2}
\rho(e_\mu^A)\,=\,
\exp\bigl[-M^2 a^2\,e_\mu^A e_\mu^A\bigr],
\end{equation}
amounts to a Gaussian integral, resulting in
\begin{equation}
\label{apf}
Z\,=\,Z_0\,\int DA_\mu\,
\exp\left[-\frac 12\sum\limits_{n\ge 2}\frac{(-1)^{n-1}}{n}
\bigl(2i k\bigr)^n a^{2n-4}\int d^4x\,\mathrm{tr}\mathcal{ F}^n\right],
\end{equation}
where the matrix products and traces run over both pairs
of the indices of the matrix $\mathcal{F}^{AB,\mu\nu}(A)$ (\ref{F})
or, equivalently, over the multiple index
$\left\{A, \mu\right\}$. The partition function (\ref{pf})
is obviously recovered when $a \to 0$ and $16\,k^2 = 1/g^2$.

The example (\ref{apf}) may be of interest as providing the full action in the
exponent beyond the leading term. In principle, it could be done for other cases,
resulting in a sum over a variety of structures made of powers of the
strength tensor $G_{\mu\nu}^{AB}$ with the coefficients evaluated through the
density function,
$S(a)=\sum_k a^k S_k$, where $S_k$ stands for terms of
dimension $2n$ in mass units, $k=2n-4\ge 0$. An important question in this context
is whether the UV divergences generated by the next terms in the full action
$S(a)$ affect the continuous limit $a \to 0$. To this end let us look at
the $N$-point Green function $G_N$ given by the Feynman diagrams with $N$ external gluon
legs. Their dimension in momentum space without external gluon
propagators is $4-N$. The maximal possible divergency coming after the extra
vertices $S_{k_1},\ldots, S_{k_n}$ are inserted into the diagrams
is estimated on dimensional grounds as
$(1/a)^{4-N+k_1+\ldots +k_n}$, with the UV momentum cutoff $\Lambda \sim 1/a$.
Thus, $G_N \sim a^{N-4}$, and all potentially dangerous contributions
from the extra terms
vanish when $a \to 0$
except for $N\leq 4$.
But it is just the divergences that cancel against the counterterms
in the course of renormalization.

\section{Relation to gravity}

The expression (\ref{Zrho}) makes a connection to gravity if to
integrate it starting first from the integral over the gauge field $A_\mu$
and treating the auxiliary fields $e^A_\mu$
as frame vectors, or tetrads (vierbein) in a curved space.
The metric tensor
is constructed for this space as $g_{\mu\nu}(x)=e_\mu^A(x)e_\mu^A(x)$
and it defines the contravariant tetrads,
\begin{equation}
\label{eg}
e_\mu^A\,=\,g_{\mu\nu}e^{A,\nu},~~~~
e^{A,\mu}\,e_\mu^B\,=\,\delta^{AB},~~~~
\det g\,=\,\det(e)^2.
\end{equation}
Due to the identity
$$
\varepsilon^{\mu\nu\lambda\sigma}\varepsilon^{ABCD}
\Sigma_{\lambda\sigma}^{\,CD}
=\,4\det(e)\cdot \bigl(e^{A,\mu}e^{B,\nu}-e^{B,\mu}e^{A,\nu}\bigr)\,
\equiv\,4\det(e)\,\Sigma^{\,AB,\mu\nu}
$$
the action in the exponent (\ref{Zrho}) takes the form
\begin{equation}
\label{SeA}
ik M^2\!\int\! d^{\,4}x\,F(A)\,=\,
4i k M^2\!\int\! d^{\,4}x\,\det(e)\,
G_{\mu\nu}^{AB}(A)\Sigma^{\,AB,\mu\nu}.
\end{equation}
The expression (\ref{SeA}) is the well-known Hilbert-Palatini action
(in Euclidean space).
Varying it with respect to the gauge field, the stationary point
turns out to be the spin connection defined for the frame
vectors, $A_\mu=\omega_\mu$. The variation of the $e^{A,\mu}$
components yields the general relativity classical equations for pure
gravity \cite{Peldan}.

Since this issue is important,
we re derive it in a way
that will be convenient later on. First, we
introduce covariant derivative $\nabla_\mu$
compatible with the metric, $\nabla_\lambda g_{\mu\nu}=0$,
and acting on the tetrad as
$$
\nabla_\mu e^{A,\nu}\,=\,\omega_\mu^{CA}e^{C,\nu}.
$$
The spin connection matrix $\omega_\mu^{AB}=-\omega_\mu^{BA}$
has a standard expression in terms of the frame vectors and their
first derivatives,
\begin{eqnarray}
\label{omega}
\omega_\mu^{AB}\,&=&\,\frac 12\,e^{A,\lambda}
\bigl(\,\partial_\mu e_\lambda^B-\partial_\lambda e_\mu^B\bigr)
\,-\,\frac 12\,e^{B,\lambda}
\bigl(\,\partial_\mu e_\lambda^A-\partial_\lambda e_\mu^A\bigr)
\nonumber \\
&&+\,\frac 12\,e^{A,\lambda}e^{B,\sigma}e_\mu^C
\bigl(\,\partial_\sigma e_\lambda^C-\partial_\lambda e_\sigma^C\bigr).
\nonumber
\end{eqnarray}
It allows for the obvious identity
$$
\partial_\mu\bigl[\,\det(e)\Sigma^{\,AB,\mu\nu}A_\nu^{AB}\bigr]
-\partial_\nu\bigl[\,\det(e)\Sigma^{\,AB,\mu\nu}A_\mu^{AB}\bigr]
$$
$$
\,=\,\det(e)\nabla_\mu\bigl(\Sigma^{\,AB,\mu\nu}A_\nu^{AB}\bigr)
-\det(e)\nabla_\nu\bigl(\Sigma^{\,AB,\mu\nu}A_\mu^{AB}\bigr)
$$
$$
=\,\det(e)\bigl[A_\nu^{AB}\nabla_\mu\Sigma^{\,AB,\mu\nu}
-A_\mu^{AB}\nabla_\nu\Sigma^{\,AB,\mu\nu}
+\Sigma^{\,AB,\mu\nu}(\partial_\mu A_\nu-\partial_\nu A_\mu)^{AB}\bigr].
$$
Furthermore, we have
$$
A_\nu^{AB}\nabla_\mu\Sigma^{\,AB,\mu\nu}
-A_\mu^{AB}\nabla_\nu\Sigma^{\,AB,\mu\nu}\,=\,
2\,\bigl[\,\omega_\mu,\,A_\nu\bigr]^{AB}\Sigma^{\,AB,\mu\nu}.
$$
These two identities permit the field-strength tensor
to be recast in the form
\begin{eqnarray}
\det(e)\Sigma^{\,AB,\mu\nu}G_{\mu\nu}^{AB}(A)&\,=\,&
\partial_\mu\bigl[\,\det(e)\Sigma^{\,AB,\mu\nu}A_\nu^{AB}\bigr]
-\partial_\nu\bigl[\,\det(e)\Sigma^{\,AB,\mu\nu}A_\mu^{AB}\bigr]
\nonumber \\
&\,+\,&\det(e)\,\Sigma^{\,AB,\mu\nu}
\biggl(\bigl[A_\mu-\omega_\mu,\,A_\nu-\omega_\nu\bigr]
-[\,\omega_\mu,\,\omega_\nu\bigr]\biggr)^{AB}\nonumber
\end{eqnarray}
valid for an arbitrary field $A_\mu$.
Expressing the commutator of the two spin connection matrices
in the last line from the same
equation written for $G_{\mu\nu}(\omega)$ we reach the net result
\begin{eqnarray}
\label{Aomega}
&&\det(e)\,\Sigma^{\,AB,\mu\nu}G_{\mu\nu}^{AB}(A)
\nonumber \\
&&\,=\,
\partial_\mu\biggl[\det(e)\Sigma^{\,AB,\mu\nu}\bigl(A_\nu-\omega_\nu\bigr)^{AB}\biggr]
-\partial_\nu\biggl[\det(e)\Sigma^{\,AB,\mu\nu}\bigl(A_\mu-\omega_\mu\bigr)^{AB}\biggr]
\\
&&\,+\,\det(e)\,\Sigma^{\,AB,\mu\nu}
\biggl(\bigl[A_\mu-\omega_\mu,\,A_\nu-\omega_\nu\bigr]
+G_{\mu\nu}(\omega)\biggr)^{AB}. \nonumber
\end{eqnarray}

The identity (\ref{Aomega}) allows to evaluate the Gaussian integral
with the action (\ref{SeA}) by replacing $A_\mu \to \omega_\mu \to \overline A_\mu$
and dropping the total derivatives by assuming periodic boundary conditions.
The integrals over $\bar A_\mu(x)$
result in an extra factors $\sim [\det e(x)]^{-6}$ in
the functional measure (\ref{Derho}), after which
the integral (\ref{Zrho}) turns into
\begin{equation}
\label{Z}
Z\,=\,\int De_\mu^A\,\tilde\rho(e_\mu^A)
\exp\left\{
4i k M^2\!\int\! d^{\,4}x\,\det(e)\,
G_{\mu\nu}^{AB}(\omega)\Sigma^{\,AB,\mu\nu}
\right\},
\end{equation}
$$
\tilde\rho(e_\mu^A)(x)=\rho(e_\mu^A)(x)\det{e}(x)^{-6}.
$$
Recalling now the relation
\begin{equation}
\label{GR}
G_{\mu\nu}^{AB}(\omega)\,=\,\frac 12 R_{\lambda\sigma\,\mu\nu}(g)
\Sigma^{\,AB,\lambda\sigma},
\end{equation}
with the Riemann tensor on the-right hand side
corresponding to the metric $g_{\mu\nu}$, we get
\begin{equation}
\label{Rg}
4i k M^2\!\int\! d^{\,4}x\,\det(e)\,
G_{\mu\nu}^{AB}(\omega)\Sigma^{\,AB,\mu\nu}\,=\,
8i k M^2\!\int\! d^{\,4}x\,R\,(g)\det{e},
\end{equation}
where $R(g)$ is the scalar curvature tensor.

This expression looks like the conventional Einstein-Hilbert
action except for the space volume that properly should be positive,
$\sqrt{g}=|\det{e}|$.
To address this point we turn back to the formula (\ref{Zrho}),
which exhibits no singularities when $\det{e}$ is small.
This means that an additional cut $|\det{e}| > \epsilon$
in the integral (\ref{Zrho}) causes an effect
of order $\epsilon$, as nothing happens when $\epsilon \to 0$.
The small but finite $\epsilon$ ensures
that the $e_\mu^A$ matrices are invertible which is assumed
by the Hilbert-Palatini action (\ref{Rg}).
On the other hand, if the matrix $g_{\mu\nu}$ has an eigenvalue
of order $\epsilon^2$
the Christoffel symbol,
$$
\Gamma_{\nu\lambda}^\mu\,=\,\frac 12\,g^{\mu\sigma}\bigl(
\partial_\lambda g_{\sigma\nu} + \partial_\nu g_{\sigma\lambda}
- \partial_\sigma g_{\nu\lambda}\bigr)
$$
would generally be of order $1/\epsilon^2$,
the Ricci tensor $R_{\mu\nu}\sim 1/\epsilon^4$
and $\det{e}\,g^{\mu\nu}R_{\mu\nu}\sim 1/\epsilon^5$.
This results in a very singular action on the right-hand side
of (\ref{Rg}) in apparent conflict with the smooth
$\epsilon \to 0$ limit. The explanation is probably
that the extremely large action in the exponent makes it oscillate rapidly
and damps the integral.
The contribution of the field configurations with $\det e$ passing
through zero are therefore strongly suppressed. It is natural
to assume that the functional integral is dominated by
the separate configurations with either
$\det e > 0$ or $\det e < 0$, with their actions being
complex conjugated,
\begin{equation}
\label{Sg}
iS(g)\,=\,\pm 8\,i k M^2\!\int\! d^{\,4}x\,R\,(g)\sqrt g.
\end{equation}
(Recall that $\det g$ is always non-negative in the Euclidean
version.)
The full integral (\ref{Z}) is then given by the sum
over these two regions, or by twice
the real part of the expression
$$
Z_g\,=\,\int De_\mu^A\,\tilde\rho\bigl(e_\mu^A\bigr)e^{iS(g)}.
$$

The action (\ref{Sg})
comprises only the metric tensor, which suggests to factorizing
$g_{\mu\nu}$ out of the integral measure,
\begin{equation}
\label{measure}
d^{16}e_\mu^A\,=\,d^{\,6}O^{AB}\,d^{10}g_{\mu\nu} g^{-\frac 12},~~~
d^{10}g_{\mu\nu}\,=\,\prod\limits_{\mu \le\nu}dg_{\mu\nu}.
\end{equation}
Here $d^{\,6}O^{AB}$ stands for six angular variables parametrizing
the rotations of the $e_\mu^A$ components over the color indices $A$.
If the density
$\rho(e_\mu^A)=\rho(e_\mu^A e_\mu^A)=\rho(g_{\mu\nu})$
depends on the metric tensor only,
as is the case for the densities (\ref{rho1}), (\ref{rho2}),
the angular integrals turn into unity due to the normalization
chosen as $\int d^{\,6}O=1$,
and one is left
with the integral written completely in terms of
the metric tensor,
\begin{equation}
\label{Zg}
Z_g\,=\,\int Dg_{\mu\nu}\,\rho(g_{\mu\nu})g^{-\frac 72}e^{\,iS(g)}.
\end{equation}
The factor $g^{-\frac 72}$ arises here from the Jacobian
in the expression (\ref{measure}) combined with the $g^{-3}$ factor
in $\tilde\rho$ density.

In fact, one can deal with a more general situation
without requiring a particular density form.
Indeed, the metric tensor and the functional measure
are invariant under $SO(4)$ gauge transformation
$$
e_\mu^A(x)\,\to\,O^{AB}(x)e_\mu^B(x),
$$
which is why the integral
$$
Z_g(O)\,=\,\int \prod\limits_{x} de_\mu^A(x)
\tilde\rho\bigl(O^{AB}(x)e_\mu^B(x)\bigr)e^{\,iS(g)}
$$
stays constant for any $O(x)$ matrices put into it, $Z_g(O)=Z_g$.
Thus it remains unchanged after gauge averaging,
\begin{equation}
\label{ZgO}
\int \prod\limits_{x}d^{\,6}O(x)\,Z_g(O)\,=\,Z_g.
\end{equation}
Due to the angular measure invariance
to $SO(4)$ rotations, $d^{\,6}(O(x)\cdot R^{-1})=d^{\,6}O(x)$,
each term averaged in the product (\ref{ZgO}),
\begin{equation}
\label{PhiO}
\Phi\bigl(e_\mu^A(x)\bigr)\,=\,\int d^{\,6}O(x)
\tilde\rho\bigl(O^{AB}(x)e_\mu^B(x)\bigr),
\end{equation}
comes out to be locally invariant,
$$
\Phi\bigl(R^{AB}e_\mu^B(x)\bigr)\,=\,\Phi\bigl(e_\mu^A(x)\bigr).
$$
There are no local invariants made of
the $e_\mu^A(x)$ components without derivatives except
for the scalar product $e_\mu^A(x)e_\nu^A(x)$ and $\det e_\mu^A(x)$.
For this reason we have
$\Phi\bigl(e_\mu^A(x)\bigr)=\Phi\bigl(g_{\mu\nu}(x)\bigr)$,
which turns the more general case back to the formula (\ref{Zg})
with a certain function $\rho(g_{\mu\nu})$ but
its relation to the input density $\tilde\rho(g_{\mu\nu})$
is not straightforward.

One can go further and work out the formula (\ref{Zg})
in a similar manner. The action $S(g)$ (\ref{Sg}) is invariant
under the coordinate transformations, or diffeomorphisms,
\begin{eqnarray}
x^\mu\,&\to &\,\,\xi^\mu(x),~~~~~~
g_{\mu\nu}(x)\,\to\,g_{\mu\nu}^\xi(x)
\,=\,g_{\lambda\sigma}(\xi(x))
\frac{\partial \xi^\lambda}{\partial x^\mu}
\frac{\partial \xi^\sigma}{\partial x^\nu},
\nonumber \\
S(g)\,&=&\,S(g^\xi).
\nonumber
\end{eqnarray}
The invariance of
the relevant gravity measure $Dg$ is achieved
by including a local factor \cite{KP},
\begin{equation}
\label{Dg}
Dg\,=\,\prod_x\prod_{\mu\le\nu}g^{\frac 52}\,dg^{\,\mu\nu}\,=\,
\prod_x\prod_{\mu\le\nu}g^{-\frac 52}\,dg_{\mu\nu},~~~
Dg\,=\,Dg^\xi.
\end{equation}
Written in terms of the invariant measure the integral (\ref{Zg}),
$$
Z_g\,=\,\int Dg \prod\limits_{x}
\rho\bigl(g_{\mu\nu}(x)\bigr)g^{-1}(x)\,e^{iS(g)}\,=\,
\int Dg\,e^{\,S_\rho+iS(g)},
$$
amounts to an extra part added to the action,
\begin{equation}
\label{Srho}
S_\rho\,=\,\frac{1}{a^4}\int d^{\,4}x\,
\ln \bigl[\rho\bigl(g_{\mu\nu}(x)\bigr)g^{-1}(x)\bigr]\,
\equiv\,\int d^{\,4}x\,\sqrt{g(x)}\,
\varphi_\rho \bigl(g_{\mu\nu}(x)\bigr).
\end{equation}
It explicitly destroys the general covariance,
$\varphi_\rho(g_{\mu\nu}^\xi) \neq \varphi_\rho(g_{\mu\nu})$.
The replacement $\varphi_\rho(g_{\mu\nu})\to \varphi_\rho(g_{\mu\nu}^\xi)$,
however, does not affect the integral,
\begin{eqnarray}
Z_g(\xi)\,&=&\,\int Dg \exp\int d^{\,4}x\,\sqrt{g(x)}
\bigl(\varphi_\rho (g_{\mu\nu}^\xi)+i S(g)\bigr)
\,=\,\int Dg \exp\int d^{\,4}x\,\sqrt{g(x)}
\bigl(\varphi_\rho (g_{\mu\nu}^\xi)+i S(g^\xi)\bigr)
\nonumber \\
&=&\,\int Dg^\xi \exp\int d^{\,4}\xi\,\sqrt{g(\xi)}
\bigl(\varphi_\rho (g_{\mu\nu}^\xi)+i S(g^\xi)\bigr)\,
=\,Z_g.
\nonumber
\end{eqnarray}
Therefore, averaging the additional term
over all diffeomorphisms in $Z_g$
results only into an overall
normalization proportional to the ''volume''
of the diffeomorphism group.
Denoting the averaging symbolically as an integral
over this group,
\begin{eqnarray}
\label{Phi}
\Phi_\rho (g_{\mu\nu})\,&=&\,
\int D\xi \exp\int d^{\,4}x\,\sqrt{g(x)}
\bigl(\varphi_\rho \bigl(g_{\lambda\sigma}(\xi(x))
\frac{\partial \xi^\lambda}{\partial x^\mu}
\frac{\partial \xi^\sigma}{\partial x^\nu}\bigr)\,
\\
&=&\,
\int D\xi \exp\int d^{\,4}\xi\,\sqrt{g(\xi)}
\bigl(\varphi_\rho \bigl(g_{\lambda\sigma}(\xi)
C_\mu^\lambda(\xi) C_\nu^\sigma(\xi)\bigr),
\nonumber \\
C_\mu^\lambda(\xi)\,&=&\,
\frac{\partial \xi^\lambda}{\partial x^\mu}
\bigr|_{x^\mu=x^\mu(\xi)}, \nonumber
\end{eqnarray}
the result is obviously diffeomorphism invariant,
$\Phi_\rho (g_{\mu\nu}^\xi)=\Phi_\rho (g_{\mu\nu})$.
This allows for a $g$ dependence
only through invariants like $\int dx \sqrt{g}$,
$\int dx \sqrt{g}R$, $\int dx \sqrt{g}R^{\,2}$, etc.,
and there is generally an infinite number of
admissible structures.
A peculiar feature that sets the functional~(\ref{Phi})
apart is the absence of derivatives of the metric tensor,
as there are no visible sources for them to emerge from upon averaging.
The only invariant without a derivative is
the invariant volume, and hence
$\Phi_\rho (g_{\mu\nu})=\Phi_\rho\bigl(\int dx \sqrt{g}\bigr)$.
Moreover,
the matrices $C_\mu^\lambda(x)$
can be treated as independent if separated
by distances exceeding the UV cutoff $a$.
Then, for the fixed $g(x)$,
the averaging would amount to the product of the same
factors, the number of which is either proportional to the total space volume
$V_4$ or, more exactly, equal to $V_4/a^4$.
This argument
forces the function into the form
$$
\Phi_\rho (g_{\mu\nu})\,=\,e^{M^4\lambda_\rho \int d^{\,4}x \sqrt{g}}
$$
specified by a single dimensionless constant $\lambda_\rho$.

Thus, we finally connect the partition function
of pure gravity with the cosmological term,
\begin{equation}
\label{Fin1}
Z_g\,=\,
\int Dg\,\exp \int d^4x\sqrt{g}\,\bigl[M^4\lambda_\rho\,
+\,8i k M^2 R\,(g)\bigr],
\end{equation}
to the YM partition function
\begin{equation}
\label{Fin2}
Z_0\,\int DA_\mu\,\exp\left[-\frac 1{g^2}\int d^{\,4}x\,
G_{\mu\nu}^{AB}(A)\,G_{\mu\nu}^{AB}(A)\right]\,=\,Z_g\,+\,Z_g^*,
\end{equation}
with the coupling $g^2=g^2(k)$ and relative normalization
$Z_0=Z_0(Ma)$.
It is remarkable that the density function collapses here to
a single constant, although their exact relation is rather involved.

\section{Implications for YM theory}

Here we deal with the case when gravity does not emerge.

As a first step we turn back to the identities (\ref{Aomega}),
where we split the second-rank tensors into their
self- and anti-self-dual parts with respect to the color indices,
\begin{equation}
\label{selfdual}
T^{AB}\,=\,\stackrel{+~~~}{T^{AB}}\,+\,\stackrel{-~~~}{T^{AB}},~~~~
\stackrel{\pm~~~}{T^{AB}}\,=\,\frac 12\,\bigl(T^{AB}\pm
\tilde T^{AB}\bigr),~~
\tilde T^{AB}\equiv\frac 12\,\varepsilon^{ABCD}T^{CD}
\end{equation}
$$
\bigl[\,\stackrel{+}{T}_1,\,\stackrel{-}{T}_2\bigr]\,=\,0,~~~~
\stackrel{+~~~}{T_1^{AB}}\,\stackrel{-~~~}{T_2^{AB}}\,=\,0,
$$
where the identity $[\tilde T_1,\tilde T_2]=[T_1,T_2]$
is responsible for the commutator
vanishing in the second line. In fact, this amounts to a decomposition of
$SO(4)$ algebra into two $SU(2)$ algebras whose generators are made of
plus or minus components.

Substituting $A=A^\pm$ into the equality (\ref{Aomega}), we immediately get
that it holds separately for plus and minus parts of the field-strength
tensor,
\begin{eqnarray}
\label{Aomegapm}
&&\det(e)\,\Sigma^{\,AB,\mu\nu}G_{\mu\nu}^{AB}(A^\pm) \\
&&\,=\,
\partial_\mu\biggl[\det(e)\Sigma^{\,AB,\mu\nu}
\bigl(A_\nu^\pm-\omega_\nu^\pm\bigr)^{AB}\biggr]
-\partial_\nu\biggl[\det(e)\Sigma^{\,AB,\mu\nu}
\bigl(A_\mu^\pm-\omega_\mu^\pm\bigr)^{AB}\bigr]
\nonumber \\
&&\,+\,\det(e)\,\Sigma^{\,AB,\mu\nu}
\biggl(\bigl[A_\mu^\pm-\omega_\mu^\pm,\,A_\nu^\pm-\omega_\nu^\pm\bigr]
+G_{\mu\nu}(\omega^\pm)\biggr)^{AB} \nonumber
\end{eqnarray}
provided we take into account that
$G_{\mu\nu}^{AB}(A^\pm)\,=\,\stackrel{\pm~~~}{G_{\mu\nu}^{AB}}(A)$.

The second step is the basic relation we used before
for the integral (\ref{Zrho})
\begin{equation}
\label{basic}
Z_0\,\int DA_\mu\,\exp\left[-\frac 1{g^2}\int d^{\,4}x\,
G_{\mu\nu}^{AB}(A)\,G_{\mu\nu}^{AB}(A)\right]\,=\,
\int DA_\mu De_\mu^A\,\rho(e_\mu^A)\,
\exp\left[\,ik M^2\int d^4x
\,F(A)\,\right]
\end{equation}
but taken now for
$$
F(A)\,=\,\varepsilon^{\mu\nu\lambda\sigma}G_{\lambda\sigma}^{AB}(A)
\Sigma_{\mu\nu}^{AB}\,=\,
\det(e)\varepsilon^{ABCD}\Sigma^{AB,\mu\nu}G_{\mu\nu}^{CD}(A).
$$
The coupling constant in the YM action in (\ref{basic})
is related to the parameter $k$ as
$1/g^2=2/9\,k^2$ and $1/g^2=k^2$ for
the densities (\ref{rho1}) and (\ref{rho2}) respectively,
although the particular form of the density and
the explicit dependence $g^2=g^2(k)$ will not be important below.

The action of the form
$$
S_H\,=\,\frac 14\int d^{\,4}x\,\det(e)\,
\Sigma^{\,AB,\mu\nu}\bigl[\,G_{\mu\nu}^{AB}
-\frac 1{2\gamma}\varepsilon^{ABCD}G_{\mu\nu}^{CD}\,\bigl]
$$
is the generalized Hilbert-Palatini action proposed by Holst
\cite{Holst}. It gives rise to the same equation of motion
for classical gravity regardless of the value of the Immirzi
parameter $\gamma$ \cite{Immirzi}. It is this second
term in the Holst action that is only
left in $F(A)$.

Using the identities (\ref{Aomegapm}) and omitting again
the total derivatives by imposing periodicity,
we bring the action
on the right-hand side of the basic relation (\ref{basic})
to the form
\begin{eqnarray}
&&i k M^2\!\int\! d^{\,4}x\,
\det(e)\varepsilon^{ABCD}\Sigma^{AB,\mu\nu}G_{\mu\nu}^{CD}(A)
\nonumber \\
&=&\,
2 i k M^2\!\int\! d^{\,4}x\,\det(e)
\bigl[\,\stackrel{+~~~}{G_{\mu\nu}^{AB}}(A)\,-\,
\stackrel{-~~~}{G_{\mu\nu}^{AB}}(A)\,\bigr]\,
\Sigma^{\,AB,\mu\nu}
\nonumber \\
\label{van}
&=&\,2i k M^2\!\int\! d^{\,4}x\,\det(e)
\,\Sigma^{\,AB,\mu\nu}
\biggl(\bigl[A_\mu^+ -\omega_\mu^+,\,A_\nu^+ -\omega_\nu^+\bigr]
- \bigl[A_\mu^- -\omega_\mu^-,\,
A_\nu^- -\omega_\nu^-\bigr]
\\
&&\,+\,\stackrel{+}{G}_{\mu\nu}(\omega)\,
-\,\stackrel{-}{G}_{\mu\nu}(\omega)\,\biggr)^{AB}.
\nonumber
\end{eqnarray}
As is evident from
the relation (\ref{GR}) and
the cyclic identity for the curvature tensor,
$$
R_{\lambda \sigma \mu \nu}\,+\,R_{\lambda \mu \nu \sigma}
\,+\,R_{\lambda \nu \sigma \mu}\,=\,0,
$$
the last term in the expression (\ref{van}) is identically zero,
\begin{eqnarray}
\det(e)
\bigl[\,\stackrel{+~~~}{G_{\mu\nu}^{AB}}(\omega)\,-\,
\stackrel{-~~~}{G_{\mu\nu}^{AB}}(\omega)\,\bigr]\,
\Sigma^{\,AB,\mu\nu} &&
\nonumber \\
=\,\det(e)\Sigma^{\,AB,\mu\nu}\varepsilon^{ABCD}{G}_{\mu\nu}^{CD}(\omega)
\,&=&\,\frac 12\det(e)\Sigma^{\,AB,\mu\nu}\varepsilon^{ABCD}
R_{\lambda \sigma \mu \nu}(g)
\Sigma^{\,CD,\lambda \sigma}
\nonumber \\
\label{van1}
\,=\,\frac 12\varepsilon^{\mu\nu\alpha\beta}\Sigma_{\alpha\beta}^{CD}
R_{\lambda \sigma \mu \nu}(g)\Sigma^{\,CD,\lambda \sigma}\,&=&\,
2\,\varepsilon^{\lambda\sigma\mu\nu}R_{\lambda \sigma \mu \nu}(g)\,=\,0.
\end{eqnarray}
It is just the vanishing of the gravity that was mentioned above.

The replacement $A_\mu \to \omega_\mu \to \overline A_\mu$ removes
any terms with derivatives from the expression (\ref{van}), turning the partition
function into the product of independent integrals,
\begin{eqnarray}
Z\,&=&\,\prod\limits_x
\int dA_\mu(x) de_\mu^A(x)\,\rho\bigl(e_\mu^A(x)\bigr)
\exp\biggl[
\,i k a^4 M^2\,\det\bigl(e(x)\bigr) \Sigma^{\,AB,\mu\nu}(x)
\bigl(\,\bigl[A_\mu^+(x),\,A_\nu^+(x) \bigr]
\nonumber \\
\label{uncorr}
&&- \bigl[A_\mu^-(x),\,A_\nu^-(x) \bigr]\,\bigr)^{AB}\biggr].
\end{eqnarray}
This result looks like
averaging over an ensemble of independent random variables.
The situation is more complex when going over to correlators.
Let us consider a specific example of the Wilson loop,
$$
W(A)\,=\,\mathrm{Tr P}\exp\oint dx^\mu A_\mu(x),
$$
where the path-ordered exponent is taken along a closed contour $x^\mu(t)$.
Applying Eq.~(\ref{basic}) to the average value yields
\begin{eqnarray}
\label{W}
\langle W\rangle\,&=&\,\int DA\,\exp\left[-\frac 1{g^2}
\int d^{\,4}x\,
G_{\mu\nu}^{AB}(A)\,G_{\mu\nu}^{AB}(A)\right]\,W(A)\\
\,&=&\,
\int DA_\mu De_\mu^A\,\rho(e_\mu^A)\,
\exp\left[\,ik M^2\int d^{\,4}x
\det(e) \Sigma^{\,AB,\mu\nu}
\left(\bigl[A_\mu^+,\,A_\nu^+ \bigr]
- \bigl[A_\mu^-,\,A_\nu^- \bigr]\right)^{AB}\right]\,
\nonumber \\
&&\times
\mathrm{Tr P}\exp\oint dx^\mu \bigl(A_\mu(x)+\omega_\mu(x)\bigr).
\nonumber
\end{eqnarray}
Now the spin connections survive in this expression, preventing
it from decaying into a product like (\ref{uncorr}).
The discretized version of the derivative implies the shifted tetrad
$e_\mu^A(x+a)$ in $\omega_\mu^{AB}(x)$ Eq.(\ref{omega}), which leads to the overlap
with the integrals at neighboring points.

Nevertheless the formula (\ref{W}) can provide some insight
into the Wilson loop behavior. Suppose the contour $x^\mu(t)$ is flat and
lies, say, in the $x^3=x^4=0$ plane.
The derivatives in the spin connection give rise to correlations
that spread in all directions around the contour over a distance
of order $a$.
They do not reach points far away
from the plane where the contour lies.
The contribution of the distant points to the $\langle W\rangle$
value (\ref{W}) is again given by the product
(\ref{uncorr}) but taken for $|x^3|,|x^4| \gg a$.
Actually it contributes merely to an overall normalization
of $\langle W\rangle$.

The integral (\ref{W}) assumes periodic boundary conditions
with the period $L \to \infty$ needed
to drop out the total derivatives in the equalities (\ref{Aomegapm}).
The above reasoning tells that if another period $T \gg a$
were chosen for the orthogonal coordinates $x^{3,4}$,
it would affect the Wilson loop only through the normalization.
Indeed, when $T$ is changed it alters the number of points
in the product (\ref{uncorr}) but has no impact on what goes on
near the $x^{3,4}=0$ plane.
Once this property is established for the right-hand side
of the relation (\ref{W}) we turn back to work out its left-hand side,
that is, to evaluate the Wilson loop in the same YM theory but
with suitable adjusted periods $T$.

A finite $T$ means
that the gluons' orthogonal momenta
take on discreet values,
$p_n^{3,4}=n^{3,4}\mu$,
for integer $n^{3,4}$, $\mu=2\pi/T$.
The momenta relevant to the Wilson loop calculation
have only flat components $p^{1,2} \sim 1/R$, where $R$
is the typical loop size.
If we chose $\mu \gg 1/R$ they would be negligible
unless $n^{3,4}=0$.
Speaking in the perturbation theory language,
the most singular in the IR region diagrams are those
where all the momenta in the propagators
are flat, that is taken for $n^{3,4} = 0$.
Thus, the IR behavior is naturally described in terms
of 2D fields $A_\mu(x_1,x_2)$, $\mu=1,2,3,4$.
The dynamics is then ruled by an effective action
\begin{eqnarray}
\label{2D}
S_{eff}(A)\,&=&\,-\frac{(2\pi)^2}{g^2\mu^2}\,
\int d^2x\,G_{\mu\nu}^{AB}G_{\mu\nu}^{AB},
~~~\mu,\nu=1,2,3,4 \\
&=&\,-\frac{(2\pi)^2}{g^2\mu^2}\,
\int d^2x\,\bigl[G_{\alpha\beta}^{AB}G_{\alpha\beta}^{AB}
+2\bigl(D_\alpha(A)\phi_k\bigr)^{AB}\bigl(D_\alpha(A)\phi_k\bigr)^{AB}
+[\phi_i,\phi_k]^{AB}[\phi_i,\phi_k]^{AB}\bigr] \nonumber \\
&&~~\alpha,\beta = 1,2,~~~~i,k=1,2, \nonumber
\end{eqnarray}
where the transverse components
actually reduce
to the scalar fields in the adjoint representation,
$A_{3,4}^{AB}(x_1,x_2)=\phi_{1,2}^{AB}(x_1,x_2)$.
The parameter $\mu$ in the action (\ref{2D}) plays
the role of the UV cutoff.

The theory (\ref{2D}) is far from being trivial,
but here the most general handling will be sufficient.
As the scalar fields $\phi_{1,2}$ are not directly coupled
to the Wilson loop, they can, in principle, be integrated out
in the functional integral.
Due to the gauge invariance, what is left after would be the action
expressed through the strength tensor of the 2D gluon fields.
The long-distance asymptotics is determined by
the term containing the minimal number of derivatives,
\begin{equation}
\label{G2}
S_{2}(A)\,=\,-\frac{1}{M_*^2}\,
\int d^2x\,G_{\alpha\beta}^{AB}G_{\alpha\beta}^{AB},
~~~\alpha,\beta = 1,2,
\end{equation}
with the mass parameter $M_*^2=M_*^2\bigl(\mu,g\bigr)$.
It gives rise to the square law for the Wilson loop and,
consequently, to the linear potential,
\begin{equation}
\label{V}
V(x)=\sigma\,|\,x\,|,~~~~\sigma\,=\,\frac{C}{8}M_*^2,
\end{equation}
with the $SO(4)$ color coefficient $C=3/2$ for the fundamental
and $C=2$ for the adjoint representation.

The parameter $\mu$ plays a role similar to that of
a factorization scale separating small
and large distances.
The dynamics at small distances is not strongly influenced
by the large size periodicity. Roughly, it is almost
the conventional 4D theory with $\mu$ as an IR cutoff.
It ''microscopically'' underlies
the effective IR 2D theory
making the coupling be $\mu$ dependent,
$g=g(a,\mu)$.
Another source of $\mu$ dependence
is the IR dynamics itself,
for which $\mu$ is the UV cutoff.
The independence
of the Wilson loop from the transverse period $T$ translates
into a kind of flow equation,
$$
\mu\frac{d}{d\mu}M_*^2\bigl(\mu,g(a,\mu)\bigr)\,=\,0,
$$
governing the IR behavior of the coupling.

The above treatment can be sketched as follows.\\
1) Since Wilson loop is independent of the orthogonal periods $T$,
we evaluate it for small $T$ values.
2) The large-distance behavior for small $T$
is described by the theory that is effectively 2D.
3) It results in the asymptotically linear potential
for the Wilson loop provided the gauge symmetry is unbroken.

\section{Discussion}

The main results of this paper are summarized in
the expressions (\ref{Fin1}), (\ref{Fin2}) and (\ref{V})
with the two constants inside, $\lambda_\rho$ and $M_*^2$,
which have not really been evaluated.
Nevertheless
the very relations they enter are rather simple and
supposed to be of general validity.

There are several issues to be commented on in this context.

\noindent
1) The key idea that the derivation is based on is pointwise
integration over the frame vectors $e_\mu^A(x)$,
which implies the discretized space with the continuous limit
to follow.
Apart from this, the rest of the derivation relies
on the identities (\ref{Aomega}) and (\ref{Aomegapm}).
The first provides a link between YM theory
and gravity.
Strictly speaking, this relation
was derived assuming
the dominance of nondegenerate configurations
in the gravity functional integral as well.
Rather, the second result (\ref{V})
was based on the identity (\ref{Aomegapm}) without
any additional assumptions.
This identity completely eliminates gravity
at least as a dynamical entity,
leaving instead the ensemble
of the noninteracting random variables.
Dealing with correlators, the interaction occurs
only in a thin layer around the correlators' points whereas
the remaining bulk produces a mere normalization.
This allows to remove unessential
degrees of freedom by imposing the period $T$
for the transverse directions,
making it similar to Kaluza-Klein models.
Choosing $T$ much smaller than the Wilson loop size,
the massive Kaluza-Klein modes get large masses $m\sim 1/T$
and split off the long-range dynamics. Still, they give virtual
corrections to soft vertices that amount to the effective
purely 2D action.
Unless the gauge symmetry is unbroken,
it will result in the linear potential
irrespective of the effective action details.

\noindent
2) The cutoff $a$ is basically implemented
in this approach, making it somewhat similar to a lattice
theory.
However the correct way, probably,
is to treat it as a method to get an IR asymptotics
of YM theory.
Then, $a$ is a typical scale separating
short and large distances.
The dynamics at short distance is
mainly perturbative in the asymptotic freedom region.
The boundary of this region could be placed at
the point where the coupling $g(a)\simeq 1$,
where here the value $a$ plays the role of an IR cutoff,
turning into a UV cutoff for the IR theory at large distances.
More precisely, one has to construct an effective
action in the Wilson sense by integrating out the fields
with large momenta $p>1/a$.
The coupling $g=g(a)$ appears in it as the output
of the perturbative Gell-Mann--Low equation.
The effective action is to then be taken as the input
for the gauge-gravity manipulations at large distances, so that
the parameters $k$ and $\lambda_\rho$ in the relations
(\ref{Fin1})--(\ref{Fin2}) are functions of $g(a)$.
Given the IR interpretation these relations
suggest that the IR limit of $SO(4)$ YM
theory looks like pure gravity,
at least for the partition function.
It fits the confinement picture in the sense
that only colorless degrees of freedom like the metric tensor
are relevant at large distances.

\noindent
3) Remarkably, $SO(4)$ theory admits what seems to be
two different descriptions of its IR limit.
Apart from a pure gravity on the one hand
there appears a kind of pure chaos (\ref{uncorr})
on the other.
The interplay between these two approaches may be
of interest in its own right.
A possible analogy could be a random walk
or Brownian motion. The ensemble of uncorrelated steps
gives rise to a true dynamics governed by the diffusion equation
or by the Schr\"odinger one after passing to imaginary time.
From this point of view, it looks like the gravity emerges in chaos.

\noindent
4) Relations
like those between the integrals (\ref{Ze}) and (\ref{pf})
can be obtained without restriction to a particular space
dimension or a gauge group. However,
the link to gravity is achieved by turning
the action (\ref{F}) into the Palatini-Hilbert one
that requires the "square" frame vectors with an equal number
of upper and lower indices.
With the "square" $e_\mu^A$ one can repeat
all of the steps leading to the gauge-gravity relations of the type
(\ref{Fin1})-- (\ref{Fin2}). For instance, they could be derived
for the 3D $SU(2)$ gauge theory.
The relation to 3D gravity in this context was addressed
in Refs. \cite{Lunev_1, Lunev_2, Ganor:1995em, Anishetty:1992xa,
Anishetty:1998xn, Diakonov:2001xg}.
The integrals over the frame vectors
are Gaussian in the 3D case but become non-Gaussian
when going over to higher dimensions.
Instead, they have been evaluated here
by pointwise integrations but at the cost of introducing a spacing $a$,
which could be justified in the IR limit.

Another point is the local noncovariant terms added to the gravity
action like that in (\ref{Srho}). They arise in any dimension,
in particular in 3D, and
were dubbed "aether" in Ref.~\cite{Diakonov:2001xg}.
Here we argue that, regardless of their form, they turn into
the cosmological term.

The "nonsquare" $e_\mu^A$ would result in various extensions
of the gravity; see the discussion in Ref.~\cite{Diakonov:2001xg}.

It is worth emphasizing that the link to chaos as another
face of the IR limit is provided by the identity (\ref{van}).
It requires the $SO(4)$ group and, consequently, 4D space,
which sets this theory apart within the approach pursued here .

\noindent
5) An important criterion
of confinement is provided by the Polyakov loop,
$$
P(x)=\mathrm{Tr P}\exp\int_0^\beta d\tau  A_0(\tau,x),
$$
where the integral is taken along the Euclidean time
direction up to the inverse temperature $\beta$.
It vanishes when the system is confined and thereby may serve
as an order parameter for the confined or deconfined phases.
On the other hand, it vanishes due to the central symmetry
of YM theory
and develops a nonzero value only if this symmetry
is spontaneously broken (see \cite{Pasechnik:2021ncb,
Reinosa:2024njc} and references therein).
One has to point out that the approach proposed here
to the Wilson loop cannot be just taken over to the Polyakov loop
for the following reason.
The averaging done in the original YM theory
differs from that in the "deformed" theory (\ref{2D})
in the normalization factors (\ref{uncorr}).
They depend on the periods $L$ and $T$ but not
on the size of the Wilson loop or, more generally, on its
shape if the loop is flat. This allows, in principle, to find
its average up to an additive constant factor in the exponent.
However this trick does not work for the Polyakov loop
since its size coincides with the period, $L=\beta$.

Essential progress was recently achieved
in the generalization of quantum field symmetries
to higher-form global symmetries, which act on
multidimensional objects
see \cite{Schafer-Nameki:2023jdn,Shao:2023gho}.
In particular, the central symmetry
is associated with a 1-form acting on one-dimensional
extended objects like a Polyakov loop, whereas the form
itself lives in the ambient space. The higher-form symmetries
are of topological nature in the sense that
their action does not change under infinitesimal space deformations.
It would be of interest if there is a way to use
this property for the trick similar to that made here for
Wilson loop.
Probably one could study the broken or unbroken central symmetry
phases by deforming the original theory.

\noindent
6) There has been much recent theoretical activity in studying
2D gauge theories with adjoint matter; see
for instance, \cite{Komargodski:2020mxz, Donahue:2019adv}.
However, they are mostly models with the adjoint Majorana fermions,
whereas adjoint scalars are rarely discussed
(see, e.g.,\cite{Mandal:2011hb}).

\noindent
7) The kind of
dimensional reduction provided here matches
the picture of confinement as being due to the formation
of the flux tube between the color charges
in the dual superconductor approach
\cite{Nambu:1975ba, Mandelstam:1974pi, tHooft:1981bkw}.
Since the tube is a spatially one-dimensional object,
the theory is effectively reduced to two dimensions.

\noindent
8) It is useful
to remark that results obtained for the $SO(4)$ group
can be translated
to $SU(2)$ theory by the fact that $SO(4)=SU(2)\otimes SU(2)$.
The decomposition (\ref{selfdual}) is easily
done in a suitable basis of $SO(4)$ generators.
Introducing the set $\tau_{\pm}^A=\{\tau^a,\mp i\}$,
where $\tau^a$ are the Pauli matrices, $a=1,2,3$,
the generators are the real antisymmetric matrices
written through the standard symbols,
$A,B = 1,2,3,4$,
$$
\tau_{+}^A \tau_{-}^B\,=\,\delta^{AB} + i\eta_a^{AB}\tau^a, ~~
\tau_{-}^A \tau_{+}^B\,=\,\delta^{AB} + i\bar\eta_a^{AB}\tau^a,~~~~
\stackrel{+~~~}{T_a^{AB}}\,=\,-\frac 12\eta_a^{AB}, ~~
\stackrel{-~~~}{T_a^{AB}}\,=\,-\frac 12\bar\eta_a^{AB},
$$
so that
$$
\bigl[\,\stackrel{\pm}{T}_a,\,\stackrel{\pm}{T}_b\,\bigr]\,=\,
\varepsilon_{abc}\stackrel{\pm}{T}_c,~~~
\bigl[\,\stackrel{+}{T}_a,\,\stackrel{-}{T}_b\,\bigr]\,=\,0,~~~
\stackrel{\pm~~~}{T_a^{AB}} \stackrel{\pm~~~}{T_b^{AB}}\,=\,\delta_{ab},~~~
\stackrel{+~~~}{T_a^{AB}} \stackrel{-~~~}{T_b^{AB}}\,=\,0.
$$
One gets in this basis
\begin{eqnarray}
A_\mu^{AB}\,&=&\,\stackrel{+~}{A_\mu^{a\vphantom{AB}}} \,\stackrel{+~~~}{T_a^{AB}}
\,+\,\stackrel{-~}{A_\mu^{a\vphantom{AB}}} \,\stackrel{-~~~}{T_a^{AB}},~~~~
G_\mu^{AB}\,=\,\stackrel{+~}{G_\mu^{\,a\vphantom{AB}}} \,\stackrel{+~~~}{T_a^{AB}}
\,+\,\stackrel{-~}{G_\mu^{\,a\vphantom{AB}}} \,\stackrel{-~~~}{T_a^{AB}},
\nonumber \\
\stackrel{\pm~~}{G_{\mu\nu}^{\,a}}\,&=&\,
\partial_\mu\! \stackrel{\pm}{A_\nu^a}
\,-\,\partial_\nu\! \stackrel{\pm}{A_\mu^a}
\,+\,\varepsilon_{abc}\!\stackrel{\pm}{A_\mu^b}
\stackrel{\pm}{A_\nu^c}. \nonumber
\end{eqnarray}
Thus, the $SO(4)$ partition function (\ref{pf}) turns into the product of
two equal $SU(2)$
partition functions for $\stackrel{\pm~}{A_\mu^{a\vphantom{AB}}}$ fields.
Similarly, the Wilson loop in the $SO(4)$ adjoint representation
can be shown to be the product of two adjoint $SU(2)$ Wilson loops
separately averaged over $\stackrel{\pm~}{A_\mu^{a\vphantom{AB}}}$ fields.


\end{document}